\documentclass[sigconf, nonacm]{acmart}
\AtBeginDocument{%
	\providecommand\BibTeX{{%
			\normalfont B\kern-0.5em{\scshape i\kern-0.25em b}\kern-0.8em\TeX}}}

\usepackage{multirow}
\usepackage{soul}
\usepackage{mathtools}
\usepackage{amsmath,amsfonts, amsthm}
\usepackage{graphicx}
\usepackage{float}
\usepackage{lipsum}
\usepackage{subfig}
\usepackage{textcomp}
\usepackage{xcolor}
\usepackage{placeins}
\usepackage{hhline}
\usepackage{tabularx}
\usepackage{enumitem}
\usepackage{setspace}
\usepackage{booktabs}
\usepackage{adjustbox}
\usepackage[normalem]{ulem}
\usepackage{indentfirst}
\usepackage{makecell}
\usepackage{titlesec}
\usepackage[T1]{fontenc}
\usepackage{textcomp, libertine}
\usepackage[most]{tcolorbox}

\usepackage{algorithm}

\usepackage[]{algpseudocode}

\makeatletter
\newcommand{\mathleft}{\@fleqntrue\@mathmargin0pt}
\newcommand{\mathcenter}{\@fleqnfalse}
\makeatother

\theoremstyle{definition}

\let\sigproof\proof\let\proof\relax
\let\sigendproof\endproof\let\endproof\relax
\let\proof\sigproof
\let\endproof\sigendproof

\makeatletter
\def\@copyrightspace{\relax}
\makeatother

\begin{document}
	\title{Enhancing Security Control Production With Generative AI}
	
    \author{Chen Ling, Mina Ghashami, Vianne Gao, Ali Torkamani, Ruslan Vaulin, Nivedita Mangam, \\ Bhavya Jain, Farhan Diwan, Malini SS, Mingrui Cheng, Shreya Tarur Kumar, Felix Candelario}
    \affiliation{%
    \institution{Amazon AWS}
    \city{New York}
    \country{United States}}
    \email{{emorycl, ghashami}@amazon.com}




    
    \renewcommand{\shortauthors}{Chen Ling et al.}

\begin{abstract}
Security controls are mechanisms or policies designed for cloud based services to reduce risk, protect information, and ensure compliance with security regulations. 
The development of security controls is traditionally a labor-intensive and time-consuming process. 
This paper explores the use of Generative AI to accelerate the generation of security controls. We specifically focus on generating Gherkin codes which are the domain-specific language used to define the behavior of security controls in a structured and understandable format. 
By leveraging large language models and in-context learning, we propose a structured framework that reduces the time required for developing security controls from 2-3 days to less than one minute. Our approach integrates detailed task descriptions, step-by-step instructions, and retrieval-augmented generation to enhance the accuracy and efficiency of the generated Gherkin code. Initial evaluations on AWS cloud services demonstrate promising results, indicating that GenAI can effectively streamline the security control development process, thus providing a robust and dynamic safeguard for cloud-based infrastructures.
\end{abstract}

\maketitle

\section{Introduction}
In today's rapidly evolving digital landscape, safeguarding the security and integrity of cloud-based infrastructures has become a critical priority. The intricate nature and vast scale of modern cloud environments, coupled with the escalating sophistication of cyber threats, necessitate the deployment of robust and dynamic security measures. At the heart of these defenses are security controls, which are specific safeguards or countermeasures designed to detect, prevent, or mitigate risks to information systems. 
The development of security controls through traditional methods involves a series of labor-intensive and time-consuming steps. Security engineers must perform detailed research to stay updated on the latest threats, vulnerabilities, and best practices. They engage in comprehensive threat modeling to identify potential risks, evaluate their likelihood and impact, and devise effective mitigation strategies. The final phase involves crafting, testing, and deploying custom code and configurations for security controls. This process is both resource and time consuming, causing delays in implementing essential security measures and leaving systems vulnerable.

\begin{figure*}[t]
\centering
\includegraphics[width=0.65\textwidth]{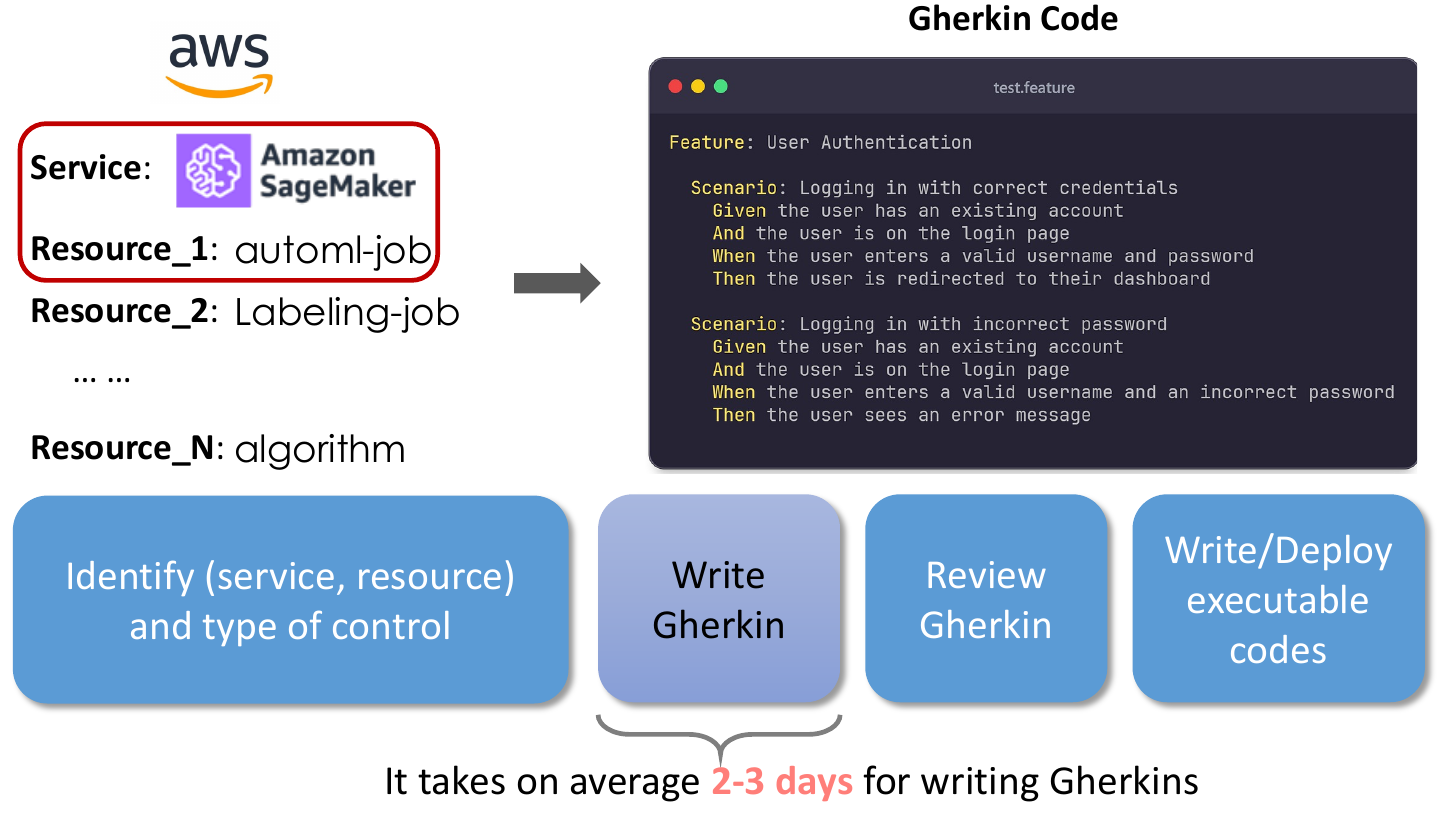} 
\vspace{-3mm}
\caption{An example of the security control development process, illustrating that the development of writing Gherkin can take on average $2$-$3$ days.}
\vspace{-5mm}
\label{fig: intro}
\end{figure*}

In light of these challenges, there is growing interest in leveraging generative AI to streamline and enhance the development of security controls. Generative AI has the potential to automate many of the labor-intensive aspects of this process, thereby significantly reducing the time and effort required to establish effective security measures. By accelerating the creation of security controls, organizations can more swiftly adapt to emerging threats, ensuring the ongoing protection and integrity of their cloud-based infrastructures.

\subsection{Development of security controls} 
The development of security controls involves multiple intricate stages, each requiring careful planning and execution to ensure the robustness and efficacy of the controls. The process begins with security engineers identifying the specific service and resource, along with the appropriate type of control required for the pair. This initial stage is crucial, as it sets the foundation for the subsequent steps by determining the scope and focus of the security measures.

Once the service, resource, and control type are identified, the next stage involves writing Gherkin scripts. Gherkin, a domain-specific language, is employed to define the behavior of security controls in a clear and structured manner. Gherkins use plain language to describe the expected outcomes, making them accessible to both technical and non-technical stakeholders. Writing Gherkin scripts requires a deep understanding of the service and resource, as well as the specific security requirements they must meet.

After the Gherkin scripts are written, they undergo a rigorous review process. This step is critical to ensure the quality and accuracy of the scripts. Security engineers meticulously review each Gherkin script to verify that it accurately defines the intended security control and that it will function correctly when implemented. 
Once the Gherkin scripts have been thoroughly reviewed and validated, the next stage is the development of the actual code to execute the control. This involves translating the Gherkin-defined behaviors into executable code that can be deployed within the cloud environment. Finally, the code is deployed, and the security controls are put into operation. This deployment stage includes thorough testing to ensure that the controls function as intended and effectively mitigate the identified risks. Continuous monitoring and maintenance are also necessary to adapt to new threats and evolving requirements.

Figure \ref{fig: intro} shows all development stages discussed above. The entire loop—from identifying and analyzing API documentation to reviewing the generated Gherkins—can span up to couple of weeks for a single service and resource pair. This protracted timeline underscores the need for innovative solutions, such as Generative AI, to automate and expedite the creation of security controls, thereby significantly reducing both the time and effort involved in their development and deployment.


\subsection{Challenges}
While Generative AI holds great promise in automating the generation of Gherkin codes, there are two practical challenges that must be addressed to fully realize its potential. 

\noindent\textbf{Challenge 1: Accurate Interpretation of Complex Service Documentation.} Cloud services are highly diverse, each with its own set of configurations, actions, and security considerations. Given the limited annotated data, the large language model needs to understand the intricacies of these services and translating them into precise and actionable Gherkin specifications without conducting finetuning. 

\noindent\textbf{Challenge 2: Ensuring Quality and Reliability of Generated Gherkin Files.} Security controls must adhere to standards to ensure they provide effective protection without introducing new vulnerabilities. This necessitates a comprehensive evaluation framework that can systematically assess the generated Gherkins for completeness and compliance with security best practices.

\subsection{Contributions} 
In this work, we propose a novel paradigm to generate Gherkins for facilitating the development of security control. To tackle Challenge 1, our process starts with a comprehensive task description for the LLM, breaking it down into detailed steps to ensure clarity and precision. Next, we provide the model with examples of existing security controls and their associated Gherkins. Finally, we present the LLM with a final query to guide it in generating the desired security control efficiently. By combining a thorough task description, illustrative examples, and a clear query, our approach enables the LLM to produce accurate and effective security controls and Gherkins, thereby reducing the time and effort required from domain experts. To address Challenge 2, we collaborate with domain experts to generate a detailed rubric to evaluate generated Gherkins from multiple dimensions, ranging from 1) whether the
generated scenarios are feasible; and 2) whether the description can correctly reflect the security control specified by the scenarios.
\section{Related Works}
\noindent\textbf{LLM for Structured Output.} As Gherkin codes are a type of highly structured output, recent advancements in LLMs \cite{ling2023domain,bai2024beyond} for structured output have demonstrated significant progress in several key areas. LLMs are being enhanced through instruction tuning and innovative approaches such as ``reflection-tuning,'' which improves the quality of training data by self-evaluation and enhancement, resulting in better output alignment \cite{li2023reflection}. Additionally, models like \textsc{GPT-4} \cite{achiam2023gpt} and \textsc{LLaMA} models \cite{touvron2023llama} have shown improvements in generating complex structured data by utilizing structure-aware fine-tuning and FormatCoT (Chain-of-Thought) techniques, significantly reducing formatting errors \cite{tang2023struc}. Furthermore, novel frameworks such as CRITIC enable LLMs to self-correct by interacting with external tools, enhancing the accuracy and reliability of their outputs \cite{gou2023critic}. These advancements highlight the potential of LLMs to handle complex, structured output tasks more effectively, paving the way for their application in diverse and sophisticated scenarios.

\noindent\textbf{Few-shot/Zero-shot LLM Prompting.} LLMs for zero/few-shot prompting has shown remarkable advancements in enhancing their reasoning and inference capabilities without requiring task-specific training examples. Notably, the introduction of strategies like Plan-and-Solve Prompting and Zero-shot Chain-of-Thought (CoT) prompting has significantly improved LLM performance on complex reasoning tasks by guiding the models to break down problems into smaller, manageable steps \cite{wang2023plan,wei2022chain,kojima2022large}. Furthermore, innovations such as UPRISE (Universal Prompt Retrieval for Improving Zero-Shot Evaluation) have improved the generalization and task adaptability of LLMs by retrieving and using relevant prompts across different models and tasks \cite{cheng2023uprise}. Additionally, methods like SelfCheck enable LLMs to verify and correct their step-by-step reasoning autonomously, thus enhancing their accuracy and reliability in zero-shot settings \cite{miao2023selfcheck}.

\noindent\textbf{Retrieval-augmented Generation (RAG).} Recent advances in RAG have significantly enhanced the capabilities of LLMs. These systems combine retrieval mechanisms with text generation models to improve performance across various tasks, including dialogue generation, machine translation, and question answering \cite{ling2023open}. Notable developments include methods like Forward-Looking Active Retrieval Augmented Generation (FLARE), which iteratively retrieves relevant information to enhance text generation accuracy \cite{jiang2023active}. Furthermore, approaches such as self-memory frameworks leverage iterative retrieval and generation to create dynamic memory pools that improve content generation \cite{cheng2024lift}. These advancements highlight the ongoing evolution and potential of RAG methods in addressing complex information needs and reducing hallucinations in generated outputs.

\section{Gherkin Generation with GenAI}

\begin{figure*}[t]
        \centering
        \includegraphics[width=0.7\textwidth]{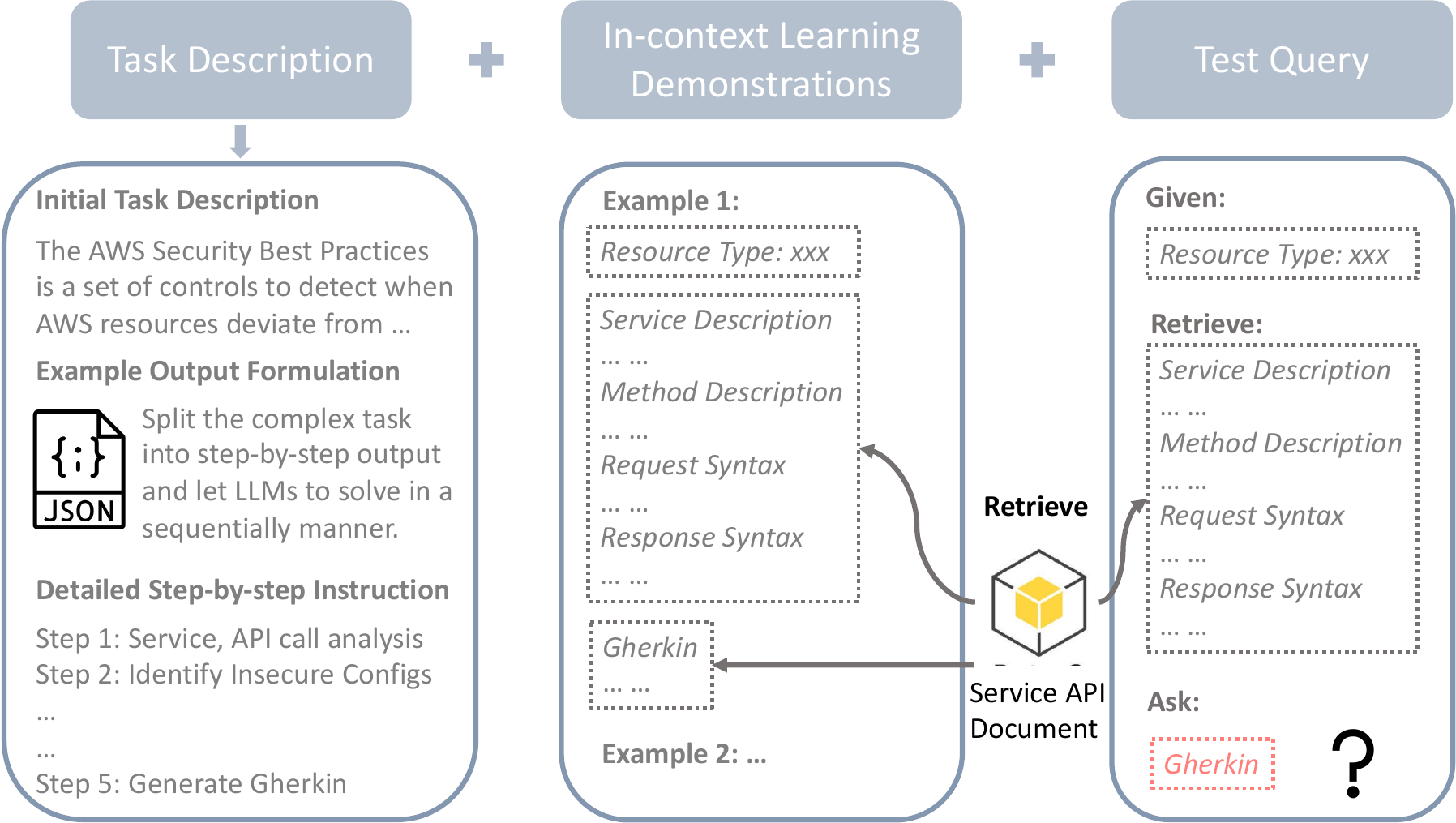}  
        \vspace{-3mm}
        \caption{An illustration of the Gherkin Generation framework.}
        \vspace{-3mm}
        \label{fig: prompt}
    \end{figure*}

\noindent\textbf{Approach Overview}. In this work, we adopt an innovative approach using in-context learning with retrieval-augmented generation to streamline the process. Our approach begins by providing the LLM with a detailed task description, breaking it down step-by-step to ensure clarity and precision. We then supply the model with existing examples of security controls, including their associated Gherkins. Finally, we present the LLM with the final query, guiding it to generate the desired security control efficiently.

\subsection{Control Types}\label{sec:control_types}
In this work, we focus on a set of critical control types identified as essential by subject matter experts (SMEs). These controls are designed to address key security and compliance requirements for cloud environments. The control types include:
\begin{itemize}[leftmargin=*]
\item \textit{Encryption of Data at Rest}: Ensuring that all stored data is encrypted to protect it from unauthorized access.
\item \textit{Encryption of Data in Transit}: Securing data during transmission to prevent interception and tampering.
\item \textit{Tagging}: Implementing consistent and meaningful tags to manage and organize resources effectively.
\item \textit{Resources Running on a Supported Version}: Ensuring that resources are running on supported versions to mitigate vulnerabilities associated with outdated software.
\item \textit{Backup Enabled}: Guaranteeing that data is regularly backed up to enable recovery in case of data loss or corruption.
\item \textit{Multi-AZ Deployment}: Distributing resources across multiple Availability Zones to enhance fault tolerance and availability.
\item \textit{Inbound IP Connection Control}: Restricting inbound IP connections to resources to prevent unauthorized access.
\item \textit{Resource Accessibility}: Ensuring that resources cannot be accessed by anyone without proper authorization.
\item \textit{Audit Logging Enabled}: Enabling audit logging and specifying the destination for logs to facilitate monitoring and compliance.
\end{itemize}
These control types have been selected based on their importance in securing cloud environments and ensuring compliance with industry standards and best practices. For a detailed description of each control type, please refer to the appendix.
\subsection{Detailed Task Description}
As seen in the left part of Figure \ref{fig: prompt}, the task description provides the LLM with an initial understanding of the overall objective of the task, which involves designing a detailed and structured prompt to guide the LLM through the generation process. The steps are as follows: 1) \textit{Task Definition}: Clearly define the task to the LLM. In this case, the task is to generate Gherkin code for a specific control type (e.g.,  logging or version support). Emphasize the role of the LLM as an expert security engineer responsible for generating these controls. 2) \textit{Context Provision}. Provide comprehensive context to the LLM about the importance of the security control. This context helps the LLM understand the rationale behind the security controls it needs to generate. 3) \textit{Expected Output.} Define the structure of the output explicitly. The output should be a JSON formatted message with no additional text. The JSON should contain specific placeholders that the LLM needs to fill with relevant information.

\subsection{Step-by-Step Instructions: CoT Reasoning}
To ensure the LLM generates accurate and effective Gherkin code for security controls, we adopt a chain-of-thought approach. This approach breaks down the task into detailed, sequential steps, facilitating clear and logical reasoning. The instructions are divided into two main steps: step 1) Service, API Call, and Insecure Configuration Analysis, and step 2) Gherkin Code Generation.

\noindent\textbf{Step 1: Service, API Call, and Insecure Configuration Analysis.} First, identify the AWS resource in question, and then identify the relevant API call that provides information about this resource. Next, understand the security best practices related to the control type (e.g., version support and management or encryption of data-at-rest). Finally, formulate a list of checks to verify compliance with security best practices. Each check should be written in a format that the LLM can interpret. These checks will help identify whether a resource is compliant with the specified security standards.

\noindent\textbf{Step 2: Generate Gherkin Code.} Based on the analysis from Step 1, the next step is to generate Gherkin code. Gherkin is a language used to write structured, human-readable tests and specifications. This code will help provide a clear and executable structure for security controls. Start by defining the rule components. This includes the Rule Identifier, which is a unique identifier for the rule, and the Rule Name, which is a descriptive name that clearly indicates the rule’s purpose. The Description should be a detailed explanation of what the rule checks for and why it is important. The Trigger specifies the condition that initiates the rule, which can be either ``Periodic'' or ``Configuration Changes''. If there are any specific parameters required for the rule, these should be defined in the Rule Parameters section. If there are no specific parameters, this section can be left empty.

\subsection{In-Context Learning and Retrieval-Augmented Generation}
After providing the detailed task description and the chain-of-thought reasoning instruction, our approach further combines the strengths of in-context learning and retrieval-augmented generation to create a robust framework for generating security controls. The process can be summarized as follows:

We design the in-context demonstrations by creating a detailed prompt that embeds examples of existing security controls to guide the LLM. We use public APIs\footnote{https://boto3.amazonaws.com/v1/documentation/api/latest/index.html} to gather background information about AWS security best practices and integrate this information into the prompt to provide the LLM with the necessary context. We then present the LLM with the final query, guiding it to generate the desired security control based on the provided context and examples. The LLM uses the embedded examples and retrieved information to produce accurate and effective Gherkin code. By leveraging these techniques, we can significantly reduce the time and effort required from security engineers to develop robust security controls. The LLM, equipped with detailed prompts and rich context, can efficiently generate Gherkin code that adheres to best practices in cloud security. This innovative approach not only streamlines the development process but also ensures the implementation of effective and reliable security measures in AWS environments.

\subsection{Agent-based Security Control Type Identifier}
\begin{figure}[t]
\centering
\includegraphics[width=0.8\columnwidth]{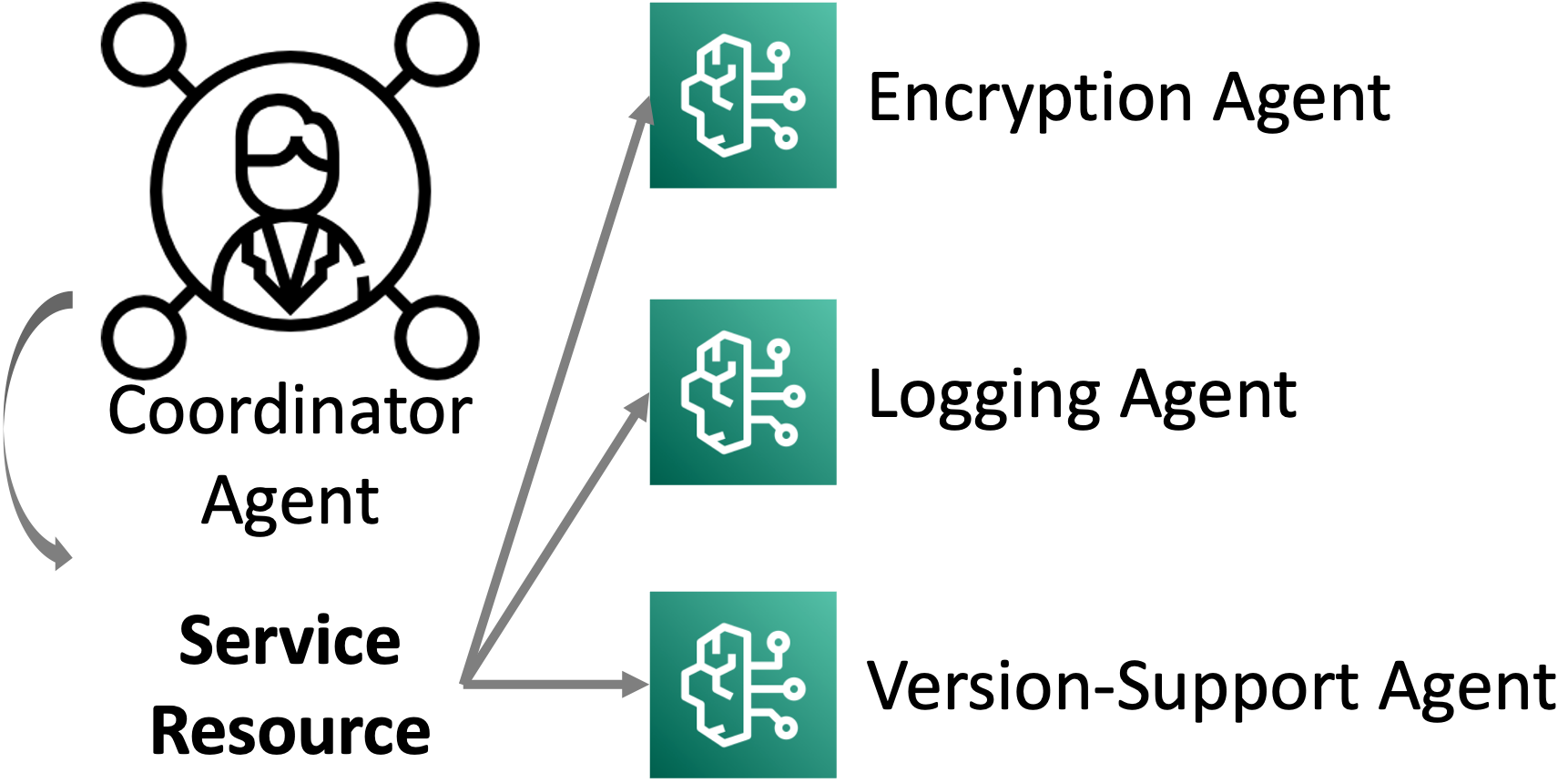} \vspace{-1mm}
\caption{The illustration of agent-based system for automated determination of control types (detailed in Section \ref{sec:control_types}) given the service and resource name.}
\vspace{-4mm}
\label{fig: parameter}
\end{figure}

As mentioned in Section \ref{sec:control_types}, security engineers at AWS have identified and categorized nine top-priority security control types for which they are interested in generating Gherkin scripts. These control types have been chosen based on their critical importance in maintaining the security and compliance of cloud environments.

In addition to the primary prompt, we have implemented a feature that allows an LLM agent to determine which security control types are applicable to a given service and resource. This feature addresses a significant challenge faced by security engineers, who often spend considerable time identifying the appropriate control types for specific cloud services and resources. The LLM agent is equipped with detailed descriptions of each Security Control Type, enabling it to make informed decisions about applicability. For instance, the control type "Encryption of Data at Rest" is defined as: "\textit{Data at rest refers to data stored in persistent, non-volatile storage for any duration. Encrypting data at rest helps protect its confidentiality, reducing the risk of unauthorized access. This control checks if the {Resource} is encrypted at rest. If the {Resource} is not encrypted at rest, the control will return NON\_COMPLIANT. If the {Resource} is encrypted, it will return COMPLIANT}."

By leveraging the LLM agent, we can automate the identification and application of relevant security controls, significantly reducing the manual effort required by security engineers. This system enhances the efficiency of the security control development process and ensures that the appropriate measures are in place to protect cloud services and resources.

\section{Evaluation System}\label{rec: eva}
To ensure the effectiveness and accuracy of Gherkin scripts generated by generative AI, we employ a human-in-the-loop approach for evaluation. This approach leverages expert human judgment to assess the quality of generated Gherkins against a structured rubric. The rubric, shown in Table 1, provides a quantitative framework for evaluating Gherkins based on specific criteria, ensuring a consistent and objective assessment. We provide a more detailed description of each criteria as follows.

\begin{table*}[h!]
    \centering
    \resizebox{0.7\textwidth}{!}{%
    \begin{tabular}{ll}
        \toprule
        \textbf{Criteria} & \textbf{Evaluation} \\
        \midrule
        S1: & The number of scenarios recorded is correct. \\
        S2: & The field specified in the scenario exists. \\
        S3: & The resulting compliance status is possible. \\
        S4: & The configuration of the resource specified by the scenario is possible. \\
        S5: & The conclusion of the scenario is correct. \\
        \midrule
        R1: & The rule name correctly describes the control specified by the collection of scenarios. \\
        R2: & The description correctly describes the control specified by the collection of scenarios. \\
        \bottomrule
    \end{tabular}%
    }
    \caption{Gherkin Rubric evaluates the validity of the generated Gherkin code from two dimensions: 1) whether the generated scenarios are feasible; and 2) whether the Rule Identifier and Description can correctly reflect the control specified by the scenarios.}
    \label{tab:gherkin-rubric}
\end{table*}

\subsection{Scenario Evaluation (S)}
\begin{enumerate}[leftmargin=*]
\item \textit{(S1) The number of scenarios recorded is correct}. This criterion assesses whether the generated Gherkin includes the appropriate number of scenarios. Each scenario should represent a distinct and necessary test case for the security control.
\item \textit{(S2) The field specified in the scenario exists}. This checks if all fields referenced in the scenarios are valid and present in the context of the security control being defined. It ensures the relevance and applicability of the scenarios.
\item \textit{(S3) The resulting compliance status is possible}. This criterion evaluates whether the compliance status derived from the scenario is feasible. It ensures that the scenarios result in legitimate compliance outcomes.
\item \textit{(S4) The configuration of the resource specified by the scenario is possible}. This ensures that the configuration actions described in the scenarios can actually be implemented within the given cloud environment. It checks for the practicality of the scenarios.
\item \textit{(S5) The conclusion of the scenario is correct}. This criterion checks if the scenario logically concludes with the correct outcome based on the preceding steps. It verifies the logical flow and correctness of the scenario's outcome.
\end{enumerate}

\subsection{Rule Evaluation (R)}
\begin{enumerate}[leftmargin=*]
\item  (R1) The rule name correctly describes the control specified by the collection of scenarios: This assesses the accuracy and appropriateness of the rule name. The rule name should succinctly and accurately reflect the control described by the scenarios.
\item (R2) The description correctly describes the control specified by the collection of scenarios: This criterion evaluates the clarity and correctness of the rule description, and whether it provides a clear understanding of the control and its purpose.
\end{enumerate}

\subsection{Final Score}
The overall score for a Gherkin script is calculated using the following formula:
\begin{equation}\label{eq: score}
    score = (S1+S2+S3+S4+S5)\times (R1+R2)/2
\end{equation}
This formula integrates the evaluations from both the scenario and rule criteria, and the final score is within the range of $[0, 5]$. The evaluation from both aspects ensures a comprehensive assessment of the Gherkin script's quality. The acceptance threshold is $\ge 2.5$ for the generated Gherkins, which indicates security engineers would need light supervision/revision to make the generated Gherkins into production.

\section{Experiments}
The experiments are conducted in collaboration with domain experts in AWS based on the evaluation metric as identified in Section \ref{rec: eva}. We use data from all available AWS services and resources. For the use of LLM, we leverage \textsc{Claude-3-Sonnet} hosted on AWS Bedrock\footnote{https://aws.amazon.com/bedrock/}.

We first demonstrate two histograms of the evaluated Gherkins by domain experts. As can be seen from Figure \ref{fig: hist}, both histograms show that the majority of the generated Gherkins for Encryption and Logging types fall within the acceptable range of requiring slight to moderate revisions (score $\ge 2.5$). The distribution patterns are similar, with most scores clustering around 3, suggesting that the generation process is relatively consistent in its output for both types. However, the consistency and concentration around the score of 3 indicate that further refinement in the generation process could help in reducing the amount of necessary revision.

We further break down the average score for two categories of security control (i.e., Encryption of data-at-rest and Logging). For each category, security engineers randomly picked ten generated Gherkins and reviewed them. The scores are depicted in Table \ref{tab:my-table}. Note that if the Final Score is greater than $4$, then the generated Gherkin can be sent to development with little modifications. If the final score is above $2.5$, the Gherkin would need moderate to slight amount of revision.

\begin{table}[t]
\centering
\resizebox{0.88\columnwidth}{!}{%
\begin{tabular}{@{}lcc@{}}
\toprule
               & \textbf{Encryption of Data-at-rest} & \textbf{Logging} \\ \midrule
Scenario Score & 4.19                       & 4.07    \\
Rule Score     & 0.72                       & 0.75    \\
Final Score    & 3.02                       & 3.05    \\ \bottomrule
\end{tabular}%
}
\caption{The average score of generated Gherkin Evaluation}
\vspace{-4mm}
\label{tab:my-table}
\end{table}
The table shows that the generated Gherkin scenarios for \textit{Encryption of Data-at-Rest} and \textit{Logging} are fairly accurate, with Scenario Scores of 4.19 and 4.07, respectively. However, the Rule Scores are low (0.72 and 0.75), indicating improvements are needed. Finally, the final score calculated by Eq. \ref{eq: score} of both control types are $3.42$, and $3.32$, respectively.

\section{Conclusion}
This study presents a novel framework for speeding up the generation of security controls using Generative AI, with a specific focus on producing Gherkin scripts. By incorporating large language models, detailed task descriptions, chain-of-thought reasoning, and retrieval-augmented generation, our approach can speed up the labor-intensive nature of traditional security control development from several days to less than one minute. Moreover, the evaluation results indicate that our method can meet the high standard of security controls evaluated by domain experts, which can be sent to deployment with minor revisions as indicated in Table \ref{tab:my-table}. In summary, this framework significantly reduces the time and effort required from security engineers while maintaining the quality and reliability of the generated controls. This advancement highlights the potential of Generative AI to enhance the development and deployment of security measures in cloud environments, providing improved protection and adaptability against evolving cyber threats.

\bibliography{cikm}
\bibliographystyle{ACM-Reference-Format}

\end{document}